\documentclass[11pt]{article}
\usepackage{amssymb}

\usepackage{pst-all}
\usepackage{pstricks}
\usepackage{pstcol,pst-fill,pst-grad}
\usepackage{amsfonts}
\usepackage{graphicx}
\usepackage{amsmath}
\usepackage[normalem]{ulem}
\usepackage{multicol}
\usepackage{tikz}
\usepackage{enumerate}

\usetikzlibrary{matrix}
\RequirePackage{mathrsfs} \RequirePackage[sc]{mathpazo}
\RequirePackage{wasysym} \RequirePackage{setspace}
\makeatletter \@addtoreset{equation}{section}

\newcommand{\be}{\begin{equation}}
\newcommand{\ee}{\end{equation}}
\newcommand{\bea}{\begin{eqnarray}}
\newcommand{\eea}{\end{eqnarray}}
\input{tcilatex}

\begin{document}

\date{}
\title{ D-brane Standard Model-Like and Scalar Dark Matter in Type IIA
Superstring Theory}
\author{Adil Belhaj$^{1}$\thanks{%
belhaj@unizar.es}, Karim Douhou$^{2}$\thanks{%
k.douhou@gmail.com}, Salah Eddine Ennadifi$^{2}$\thanks{%
ennadifis@gmail.com} \\
\\
{\small $^{1}$}{LIRST, Facult\'{e} Polydisciplinaire, Universit\'{e}
Sultan Moulay Slimane, B\'{e}ni Mellal, Morocco } \\
{\small $^{2}$}{LHEP-MS, Facult\'{e} des sciences de Rabat, Universit\'{e} Mohammed V, Rabat, Morocco} \\
}
\maketitle

\begin{abstract}
In light of the present LHC Run II at $\sqrt{s}=13$ $TeV$, string y
standard-like model is studied. Concretely, a singlet $S $ scalar-extended
SM given in terms four stacks of intersecting D6-branes in a type IIA
superstring compactification producing a large gauge symmetry is examined.
The involved scales are dealt with. According to the dark matter relic
density, the mass of the scalar dark matter beyond the SM $m_{S}\lesssim
10^{3}GeV$ and the corresponding Higgs portal couplings $\lambda
_{SH}\lesssim 10^{-8}$ are approached. \newline
\newline
\textbf{Keys words }\textit{: LHC; Standard Model; D-brane; String Theory;
Dark Matter.}\newline
\textbf{PACS}\textit{: 12.10.-g, 11.25.Uv, 12.60.Jv, 12.10.Dm}.
\end{abstract}


\thispagestyle{empty}

\newpage \setcounter{page}{1} \newpage

\section{Introduction}

More recently, many excesses with beyond the Electroweak scale $\sim
10^{2}GeV$ from LHC Run-II with $pp$ collisions at 13 $TeV$ have been
reported \cite{1,2}. These events have received a huge interest exploring
different approaches and methods using analytical and simulating studies.
These methods have been extensively investigated to provide possible
physical interpretations of such problems. Concretely, various attempts have
been suggested using models relying on extensions of the Standard Model of
Particle Physics (SM) \cite{3,4,40,5,6,7,700}. In this way, the important
investigation is based on the incorporation of singlet scalars to SM sector.
In particular, the corresponding physics could be associated with a scalar
field $S$ with a mass beyond $10^{2}GeV$. In the $pp$ collision, the
processes for producing such a scalar field are naturally obtained using two
possible ways based on the fusion of either the gluons
\begin{equation}
gg\rightarrow SS\rightarrow SM,  \label{eq1}
\end{equation}%
or the quarks
\begin{equation}
qq\rightarrow SS\rightarrow SM.  \label{eq2}
\end{equation}%
In these scenarios, the couplings of such a field could be then described by
the following effective terms
\begin{equation}
\mathbf{\zeta }\supset \left( S/\Lambda \right) \left( G_{\mu \nu
}^{a}G^{a\mu \nu }+F_{\mu \nu }F^{\mu \nu }\right).  \label{eq3}
\end{equation}
Here, $F_{\mu \nu }$ and $G_{\mu \nu }$ are the strong and the
electromagnetic fields, respectively. These terms could predict the
associated excesses and production channels.

It has been proposed that effective field theory models can be derived using
the compcatification of type II superstrings and related models allowing one
to present a possible interpretation of such a new physics \cite%
{70,71,72,73,74,75}. In string theory, the particle physics
ingredients can be provided by intersecting D-branes wrapping non
trivial cycles in orientifold compactifications. In this discussion,
the gauge symmetries can be derived from stacks of D-branes filling
the four dimensional space-time while the matter fields reside at
their intersections. The latters are associated with intersection
numbers corresponding to restrictions of additional global U(1)'s
exhibited by the compactification scenario. In this way, the stringy
effects can produce corrections to the superpotential by including
the missing coupling terms being relevant for the fermion masses.
This feature can bring an acceptable effective low-energy
realization for SM-like and their extensions
\cite{8,9,90,91,10,11,12,13}. Such models usually are represented by
graphs encoding the gauge symmetry and matter content in terms of
vertices and edges, as in quiver dual discussions. These fundamental
pieces allow for an possible exploration of several physical
problems without the need of a physical defined model. More
precisely, the possible interaction couplings can be obtained using
quantum numbers associated with graph theory representation of
D6-brane models. This graph theory method provides a rich D-brane
discussion in type II superstring compactifications
\cite{14,15,16,17,18}.

The aim of this work is to contribute to these activities by investigating a
stringy scalar-extended SM in terms of intersecting D-brane models in type
IIA superstring compactifications. To be concrete, we build a gauge theory
from intersecting D6-branes wrapping non trivial 3-cycles in a type IIA
orientifold geometry. In particular, we consider a model with $\mbox{U(3)}%
\times \mbox{Sp(1)}\times \mbox{U(1)}\times \mbox{U(1)}$ gauge symmetry. In
the corresponding SM-like, an added singlet scalar $S$, in the presence of
the standard Higgs doublet $H$, generates the SM particle masses. According
to the known data, the VEV $\left \langle S\right \rangle $ and the mass
scale of the involved new $m_{s}$ scalar provide a probe of the stringy
physics effect in the SM and a possible scalar dark matter (DM) candidate.

The organization of this paper is follows. In section 2, we present a gauge
model from four stacks of intersecting D6-branes wrapping 3-cycles in type
IIA geometry. This compactification gives $\mbox{U(3)}\times \mbox{Sp(1)}%
\times \mbox{U(1)}\times \mbox{U(1)}$ gauge symmetry. In section 3, we
propose a stringy singlet scalar extension of the SM in terms of D6-branes
in type IIA superstring. In section 4, we approach the involved high scales
associated with the new scalar mass $m_{s}$ probing the stringy effect in
the SM scale. In section 5, we present a stringy scalar DM. The last section
is devoted to concluding remarks.

\section{D6-brane SM-Like in type IIA superstring}

Motivated by the recently LHC Run-II activities and the large emergence of
scalar moduli in string theory compactifications, we study a scalar
extension of the SM by assuming that the corresponding physics involves a
stringy origin from of a low-scale $M_{s}\ll M_{Planck}$ effect. To be
precise, we will consider a singlet scalar field under the SM gauge symmetry
originated from string theory. It is recalled that non perturbative string
theory requires the introduction of objects called D-branes providing
nonabelian gauge symmetries in the lower dimensional compacatifications.
These extended objects have been explored for phenomenological applications
in string theory framework. In fact, type II superstrings contain various
solutions of D$p$-branes considered as as $(p+1)$-dimensional subspaces on
which open strings are stretched \cite{182}. The spectrum of fluctuations of
such a physics is obtained by quantizing closed strings and open strings
living on such D$p$-branes. Indeed, quantum descriptions of such objects are
given in terms of nonabelian gauge theories in $(p+1)$-dimensional physical
spaces. The corresponding physics has been extensively studied in order to
look for models close to the reality. It has been suggested that such kind
of models can be embedded in type II superstring compactifications in the
presence of D-branes producing four dimensional gauge theories. In
particular, it has been learned how many non trivial gauge models are
obtained using different methods. One method is based on singular limits of
type II superstrings by exploring the geometric engineering method \cite{181}%
. In this method, the gauge symmetry and the matter can be derived from the
geometry of the internal space by wrapping D2-branes on blowing down cycles
in the K3-fibration manifolds. Another way, which will be interested in
here, is based on intersecting D$p$-branes in type II superstrings \cite{180}%
. More precisely, we thus construct a type IIA stringy model based on four
stacks of intersecting D6-branes in the presence of a flavor symmetry
distinguishing various matter fields from each others especially quarks. It
is noted, in passing, that D5-branes in type IIB could be also used. The
latter can be related D6-brane through mirror symmetry in the Calabi-Yau
manifolds. In the intersecting D6-brane representation, the studied model is
described by the following gauge symmetry
\begin{equation}
\mbox{U}(3)_{a}\times \mbox{Sp}(1)_{b}\times \mbox{U}(1)_{c}\times \mbox{U}%
(1)_{d}.  \label{eq4}
\end{equation}%
Here, the $\mbox{Sp}(1)\simeq \mbox{SU}(2)$ weak factor symmetry arises from
the D6-wrapped on an orientifold invariant 3-cycle ($b=b^{\ast }$). It has
been observed that there is no difference between the quark doublets since
they involve all the same $\mbox{U}(1)_{a,c,d}$ abelian charges. A deeper
investigation reveals that one can examine a D6-brane model from the
compactification of type IIA superstring on three factors of the the torus $%
T^{2}$ known by factorisable torus backgrounds $\Pi
_{i=1}^{3}T_{i}^{2}$. The toroidal D6-brane configuration that we
consider here is in non-supersymmetric. Indeed, even if we consider
a compactification where supersymmetry remains unbroken, RR tadpole
conditions imply  that in a chiral D-brane configuration
supersymmetry will be broken in the open string sector of the
theory. However, it is still possible to ask if some open string
subsector will preserve some amount of supersymmetry. Or,  what are
the D-brane conditions  preserving   a common supersymmetry when
wrapping a generic compact manifold. Moreover, for consistencies,
the cancelation  of potential anomalies arising from the low energy
chiral spectrum is implied by the RR tadpole conditions, and mixed
an higher anomalies are cancelled by the generalized Green-Schwarz
mechanism. As a consequence of such  a mechanism, some Abelian gauge
bosons will get massive, eliminating the corresponding  U(1) gauge
symmetry from the effective theory   being of great importance in
constructing semi-realistic models.  In this compcatification, the
intersection numbers are given in terms of the wrapping numbers of
the D6-branes around the $T^{2}$ factors. An adequate choice of such
numbers gives intersections listed in table 1.

\begin{table}[th]
\label{t:one}
\par
\begin{center}
\begin{tabular}{|c|c|c|c|c|c|c|c|}
\hline
$Sector$ & $ab$ & $ac$ & $ac^{\ast }$ & $ad$ & $ad^{\ast }$ & $db$ & $%
dc^{\ast }$ \\ \hline
$Inter\sec tion$ & $3$ & $-2$ & $-1$ & $-1$ & $-2$ & $3$ & $-3$ \\ \hline
\end{tabular}
\label{t:one}
\end{center}
\caption{Intersection numbers of the SM spectrum. The other ones are set to
zero.}
\end{table}

In this type IIA representation, the three left-handed quarks $q^{i}$ are
localized at the intersections of D6-branes $a$ and $b$ while the
right-handed quarks, $\overline{u}^{i}$ and $\overline{d}^{i}$ split into
two up quarks $\overline{u}^{2,3}$ and one down quark $\overline{d}^{3}$
being localized at intersection of the D6-branes $a$ and $c/c^{\ast }$,
respectively. Two down quarks $\overline{d}^{1,2}$ and one up quark $%
\overline{u}^{1}$ are localized at the intersection of the D6-branes
$a$ and $d/d^{\ast }$. However, the three left-handed leptons $\ell
^{i}$ appear at the intersection of D6-branes $b$ and $d/c,c^{\ast
}$, respectively. Moreover, the three right-handed electrons
$\overline{e}^{i}$ are localized at the intersection of D6-branes
$d$ and $c^{\ast }.$ Finally, the Higgs doublet $H$ appears at the
intersection of D6-branes $b$ and $c/c^{\ast }$. It has been
remarked that the matter fields are associated with a linear
combination of $\mbox{U}(1)_{a,c,d}$ recovering the SM hypercharge.
The four $\mbox{U}(1)_{a,b,c,d}$  symmetries $Q_{a}$, $Q_{b}$, $%
Q_{c}$ and $Q_{d}$ have clear interpretations in terms of known
global symmetries of the SM, i.e., baryon, lepton and isospin
numbers. Thus,  all these known global symmetries are in fact gauge
symmetries in such stringy constructions. The hypercharge is
given here by the anomaly-free linear combination $Y=\frac{1}{6}Q_{a}-\frac{1%
}{2}Q_{c}-\frac{1}{2}Q_{_{d}}$. This D6-brane model can be graphically
illustrated in the figure 1.

\begin{center}
\begin{figure}[th]
\begin{center} {\includegraphics[width=10cm]{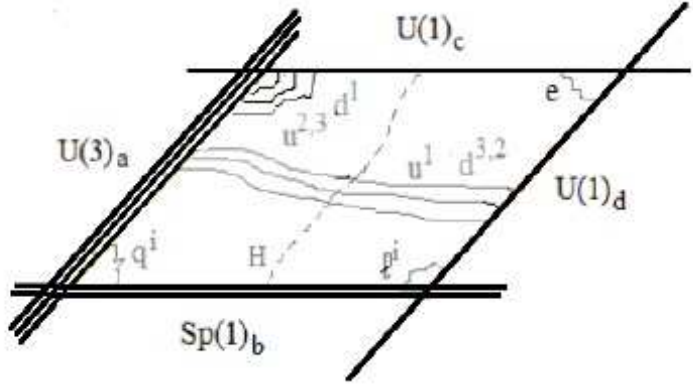}}
\end{center} \caption{4-Stack Stringy SM-Like. Bold lines denote
D6-branes and thin lines denote chiral and scalar spectrum.}
\end{figure}
\end{center}

It is observed from the figure 1 that the 4D Yukawa coupling terms can be
derived with respect to the symmetry charges illustrated in the table 2.

\begin{table}[th]
\label{t:one}
\par
\begin{center}
\begin{tabular}{|c|c|c|c|c|c|c|c|c|}
\hline
$Sector$ & $ab$ & $ac$ & $ac^{\ast }$ & $ad$ & $ad^{\ast }$ & $db$ & $%
dc^{\ast }$ & $bc$ \\ \hline
$Fields$ & $q^{i}$ & $\overline{u}^{2,3}$ & $\overline{d}^{3}$ & $\overline{u%
}^{1}$ & $\overline{d}^{1,2}$ & $\ell ^{i}$ & $\overline{e}^{i}$ & $H$ \\
\hline
$Rep$ & $3(3,\overline{2})$ & $2(\overline{3},1)$ & $1(\overline{3},1)$ & $1(%
\overline{3},1)$ & $2(\overline{3},1)$ & $3(1,\overline{2})$ & $3(1,1)$ & $%
1(1,2)$ \\ \hline
$Q_{a}$ & $1$ & $-1$ & $-1$ & $-1$ & $-1$ & $0$ & $\ 0$ & $0$ \\ \hline
$Q_{c}$ & $0$ & $1$ & $-1$ & $0$ & $0$ & $0$ & $-1$ & $1$ \\ \hline
$Q_{d}$ & $0$ & $0$ & $0$ & $1$ & $-1$ & $1$ & $-1$ & $0$ \\ \hline
$Y$ & $1/6$ & $-2/3$ & $1/3$ & $-2/3$ & $1/3$ & $-1/2$ & $1$ & $-1/2$ \\
\hline
\end{tabular}
\label{t:one}
\end{center}
\caption{SM spectrum and their U(1)$_{a,c,d}$ charges for $Y=\frac{1}{6}%
Q_{a}-\frac{1}{2}Q_{c}-\frac{1}{2}Q_{_{d}}$. The index $i=1,2,3$ is the
family index.}
\end{table}

In fact, the $\mbox{U(1)}_{a,c,d}$ field charges can be used to get the
possible Yukawa couplings associated with the interaction terms for the
heavy quarks and leptons. Computations show that these terms can be given by

\begin{eqnarray}
\mathbf{Q}_{c,d}\left( H^{\dagger }q\overline{c}\right)  &=&\mathbf{Q}%
_{c,d}\left( H\right) +\mathbf{Q}_{c,d}\left( q\right) =\mathbf{Q}%
_{c,d}\left( \overline{c}\right) =0,  \notag \\
\mathbf{Q}_{c,d}\left( H^{\dagger }q\overline{t}\right)  &=&0,\text{ }%
\mathbf{Q}_{c,d}\left( Hq\overline{b}\right) =0,\text{ }  \label{eq5} \\
\mathbf{Q}_{c,d}\left( H\ell ^{i}\overline{e}^{i}\right)  &=&0.  \notag
\end{eqnarray}%
These charges allow one to express the following Lagrangian
\begin{equation}
\mathbf{\zeta }_{Yuk}=y_{c}H^{\dagger }q\overline{c}+y_{t}H^{\dagger }q%
\overline{t}+y_{b}Hq\overline{b}+y_{e^{i}}H\ell ^{i}\overline{e}^{i},
\label{eq6}
\end{equation}%
where $y^{\prime }s$ are coupling constants associated with the
Higgs-fermion interaction strenghts between such terms.

\section{ Stringy singlet scalar extension}

An inspection in the above model reveals that the missing
phenomenologically desired coupling terms can be recovered by the
U(1)'s charged scalars which can be explored to extend the SM
spectrum \cite{14,15,16}. Besides the dilaton and the axion living
in ten dimensions, a scalar can be obtained from many roads using
the compactificaion scenario in type II superstrings. In particular,
it can be derived from the geometric deformation of the internal
space including the antisymmetric B-field of the NS-NS sector. It
has been shown that this contribution involves the complex structure
deformations, the complexified K\"{a}hler deformations or both. Or,
it comes from the R-R gauge fields on non trivial cycles of the
internal geometry in closed string sector.  In open string sector,
however,  the scalars can be obtained  from the moduli space of
deformations of special Lagranagian submanifolds, associated with
the middle-homology of the internal space, where D6-branes can wrap.
String theory compactification complexifies these scalars by adding
the Wilson lines obtained from the gauge fields living on  such
D6-branes. Here, we consider a scalar associated  with   such an
open string sector. This operation generates the missing interaction
terms with respect to the above discussed U(1) symmetry charges. A
close inspection shows that the new scalar must have the charges
listed in table 3.

\begin{table}[th]
\begin{center}
\begin{tabular}{|c|c|c|c|c|c|c|}
\hline
$Sector$ & $Field$ & $Rep$ & $Q_{a}$ & $Q_{c}$ & $Q_{d}$ & $Y$ \\ \hline
$cd$ & $S$ & $1(1,1)$ & $\ 0$ & $1$ & $-1$ & $0$ \\ \hline
\end{tabular}%
\end{center}
\caption{New scalar $S$ and its U(1)$_{c,d}$ charges.}
\end{table}

These charges can be handled to engineer a D6- brane model. The
corresponding data is represented in the figure 2.

\begin{center}
\begin{figure}[th]
\begin{center}
{\includegraphics[width=8cm]{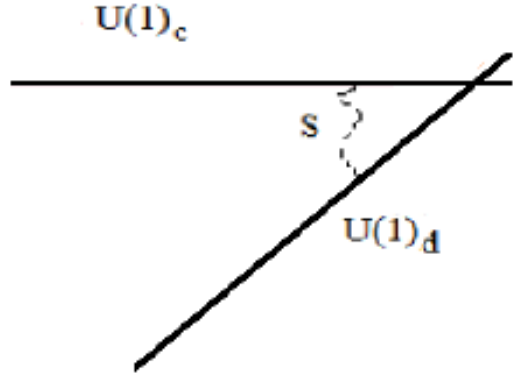}}
\end{center}
\caption{Stringy scalar and its associated U(1)$_{c,d}$ charges
indicated by the dotted line.}
\end{figure}
\end{center}

In the present D6-brane model, the absent terms are now generated from the
higher order terms. Thus, they will be suppressed by factors $\left \langle
S\right \rangle ^{n}/M_{s}^{m}$, where $n$ and $m$ are power integer
numbers. It is noted that $M_{s}$ indicates the string mass and $\left
\langle S\right \rangle $ its VEV. Indeed, these terms take the following
forms
\begin{eqnarray}
\mathbf{Q}_{c,d}\left( SH^{\dagger }q\overline{u}\right) &=&Q_{c,d}\left(
S\right) +Q_{c,d}\left( H\right) +Q_{c,d}\left( q\right) =Q_{c,d}\left(
\overline{u}\right) =0,  \notag \\
\mathbf{Q}_{c,d}\left( S^{\ast }Hq\overline{d}\right) &=&0,\text{ }\mathbf{Q}%
_{c,d}\left( S^{\ast }Hq\overline{s}\right) =0,  \label{eq7} \\
\mathbf{Q}_{c,d}\left( S^{2}(H\ell ^{i})^{2}\right) &=&0.  \notag
\end{eqnarray}
These new charges provide the following Lagrangian
\begin{eqnarray}
\mathbf{\zeta }_{Yuk}^{\prime } &=&M_{s}^{-1}\left( y_{u}SH^{\dagger }q%
\overline{u}+y_{d}S^{\ast }Hq\overline{d}+y_{s}S^{\ast }Hq\overline{s}\right)
\notag \\
&&+M_{s}^{-3}y_{\nu ^{i}}S^{2}(H\ell ^{i})^{2}.  \label{eq8}
\end{eqnarray}%
In the D6-brane representation, the VEV $\left \langle S\right
\rangle $
induces the missing Yukawa coupling terms, along with the Higgs VEV, it
generates the masses for these light fermions (\ref{eq8}). Compared to the
previous contributions given in (\ref{eq6}), they are suppressed by the
string mass scale $M_{s}{}$ with a high suppression for the left-handed
neutrino terms.

\section{Probe of the stringy scales}

It turns out that results beyond the interaction terms could be derived from
type IIA superstring moving on particular geometries. Concretely, they
include the involved string scale $M_{s}$ with the VEV $\left \langle
S\right \rangle $ and the mass $m_{s}$ of the scalar field $S$. After the
electroweak symmetry breaking by the Higgs VEV at $\left \langle H\right
\rangle \simeq 246$ $GeV$, \ appropriate combinations of fermion masses, for
which their net scalar-fermion couplings could be absorbed, give approximate
values of the new scales. In particular, using the left-handed neutrino mass
terms appearing in (\ref{eq8}) with an upper bound of $m_{\nu _{\tau
}}\lesssim 1$ $eV$, \ we can \ predict the string scale $M_{s}$. The
calculation gives the following scale
\begin{equation}
M_{s}=\frac{y_{\nu _{\tau }}}{y_{u}^{2}}\frac{m_{u}^{2}}{m_{\nu }}\sim
10^{4}GeV,  \label{eq9}
\end{equation}%
and then the scalar field VEV $\left \langle S\right \rangle $ becomes
\begin{equation}
\left \langle S\right \rangle =\frac{y_{c}}{y_{s}}M_{s}\frac{m_{s}}{m_{c}}%
\sim 10^{3}GeV.  \label{eq10}
\end{equation}%
At this level, it is worth noting that there are two new high scales: one
belongs to the low-string scale $M_{s}$ (\ref{eq9}), and the other one
belongs to the VEV of the new scalar $\left \langle S\right \rangle $ given
in eq.( \ref{eq10}). Besides the partial explanation of the fermion mass
hierarchies and the smallness of neutrino masses, these new scales appearing
in (\ref{eq9}) and (\ref{eq10}), allow a possibility to probe the type IIA
stringy effect at the SM scale through the mass of the new scalar $S$. It is
remarked that the most general renormalizable scalar potential consistent
with the model scalar spectrum reads as
\begin{equation}
V(H,S)=V(H)+\mu _{S}^{2}S^{2}+\lambda _{SS}S^{4}+\lambda _{SH}\left(
H^{\dagger }H\right) S^{2}  \label{eq11}
\end{equation}%
Form the potential, we can derive the mass of this scalar. Indeed, one read
the mass of scalar $S$ as
\begin{equation}
m_{S}=\sqrt{\mu _{S}^{2}+\lambda _{SH}v_{H}^{2}}\leqslant 10^{3}GeV.
\label{eq12}
\end{equation}%
It is noted that the constraint $\mu _{S}^{2}$ $>0$ implies the following
inequality

\begin{equation}
m_{S}>\sqrt{\lambda _{SH}}v_{H}.  \label{eq13}
\end{equation}%
At this stage, one see clearly that for the strong scalar-SM coupling value $%
\lambda _{SH}\sim 1$, the mass of the new scalar can go over the SM scale.
This high scale physics effects that result in a stringy prediction for
physics beyond the SM push to ask whether the present LHC Run II at $\sqrt{s}%
=13$ $TeV$ could be able to see more significant stringy physics directly.
Then, we should see what insights on related modern physics problems,
especially DM, can be brought.

\section{Stringy scalar Dark Matter}

The scalar singlet $S$, introduced in this type IIA D6-model, interacts with
the SM particles through the Higgs portal. It is suggested that if such a
scalar is real and stable, it could constitute a viable dark matter (DM)
candidate. Indeed, a global U(1) symmetry removing the odd power couplings
can be assumed in the simple scalar potential of the present model given in (%
\ref{eq11}) to guarantee the DM stability. It has been pointed out
that such a scalar singlet model is the simplest UV-complete theory
containing a WIMP. After the electroweak symmetry breaking, the
scalar DM candidate $S$ can be annihilated into all SM particles
through the portal coupling $\lambda _{SH}$. This operation can be
illustrated in figure 3.

\begin{center}
\begin{figure}[th]
\begin{center}
{\includegraphics[width=6cm]{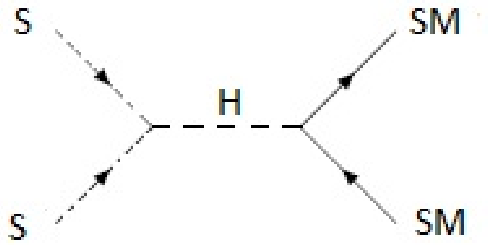}}
\end{center}
\caption{Higgs exchange diagrams for the scalar DM annihilation.}
\end{figure}
\end{center}

In the present model, there are only two relevant parameters for the DM
investigation, namely the physical DM mass $m_{S}$ (\ref{eq12}) and the
Higgs portal coupling $\lambda _{SH}$. In fact, the latter is however enough
to allow for a contribution to the invisible decay of the Higgs boson,
scattering of $S$ on nucleons through Higgs exchange, and annihilation into
SM particles. This can lead to possible indirect detection signatures and an
allowed thermal relic density of DM \cite{19,20}. In this way, the relic
density in the present universe is approximately given by the annihilation
cross-section of our scalar DM as

\begin{equation}
\Omega _{DM\equiv S}=\frac{0.1pb}{h^{2}\left \langle \sigma _{an}v_{S}\right
\rangle _{SS\rightarrow SM}}\simeq 0.2  \label{eq14}
\end{equation}%
where $h$ is the Hubble constant and where $\left \langle \sigma
_{an}v_{S}\right \rangle $ is the velocity-averaged annihilation
cross-section. Using (\ref{eq9}), (\ref{eq10}), (\ref{eq12}) and taking into
account the dilution of the scalar DM, we show that the resulting scalar DM
abundance is
\begin{equation}
\Omega _{S}\simeq 0.2\left( \frac{\lambda _{SH}}{10^{-8}}\right) ^{3}\frac{%
\left \langle S\right \rangle }{m_{S}}.  \label{eq15}
\end{equation}%
It follows for some mass range $m_{S}$ of the singlet scalar $S$ that the
natural values of $\lambda _{SH}$ reproduce not only the observed DM relic
density $\Omega _{DM}\simeq 0.2$ (\ref{eq12}), but also predict a cross
section for scattering on nucleons being not far from the current direct
detection limit. Thus, once the relic density constraints are used, one can
make definite predictions in this model as well as in different experiments
and their interplay. Given the parameter space of the model

\begin{equation}
\left \{ m_{S},\text{ }\lambda _{SH}\right \} ,  \label{eq16}
\end{equation}%
we display the range of the parameters given in the plan parameterized by $%
m_{S}$ and $\lambda _{SH}$. Moreover, we find the allowed region for the
correct relic abundance for the scalar DM satisfying the current constraint
\cite{19,20}. According to (\ref{eq10}), (\ref{eq12}), the likely mass range
of the scalar DM is
\begin{equation}
10^{2}GeV\leqslant m_{S}\leqslant 10^{3}GeV.  \label{eq17}
\end{equation}%
This data requires the Higgs portal coupling range
\begin{equation}
10^{-10}\leqslant \lambda _{SH}\leqslant 10^{-8}.  \label{eq18}
\end{equation}%
In this approach, the larger (smaller) values of $\lambda _{SH}$ reduces
(enhances) the $S$ relic density by increasing (decreasing) the annihilation
cross section. In particular, the overall predicted signal for scattering on
nucleons and the annihilation into SM particles are considered. This can be
illustrated in figure 4.

\begin{center}
\begin{figure}[th]
\begin{center}
{\includegraphics[width=6cm]{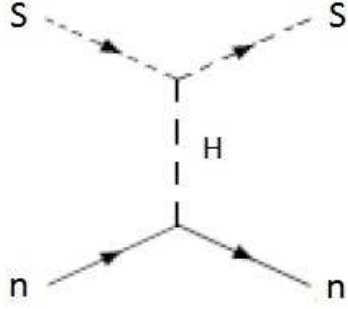}}
\end{center}
\caption{Higgs exchange diagrams for scattering with a nucleon.}
\end{figure}
\end{center}

Due to this effect, the dependence constraints are significantly different
than one might have expected. Thus, we take the view here that the singlet
scalar DM might provide only a fraction of (more than) the total DM density
which is considered as a logical possibility.

\section{Conclusion and related remarks}

In this work, we have investigated the stringy physics effect at low scales
in a string-inspired gauge theory derived from the compactification of the
type IIA superstring with D6-brane configurations. More precisely, we have
examined four stacks of intersecting D6-branes wrapped on non trivial cycles
in type IIA geometry. This model has been combined with a singlet scalar $S$
to produce an extended SM spectrum. The associated effective scalar
potential generates different coupling scales relative to the allowed
perturbative and the higher order suppressed terms. Associating the allowed
perturbative terms with the known heavy quarks and leptons, and using the
higher order generated terms corresponding to known light quarks and
neutrinos, the hierarchy of fermion masses finds a possible explanation
through higher order terms suppressed by the factors $\left \langle S\right
\rangle ^{n}/M_{s}^{m}$. Using known data, we have discussed the stringy
effect by examining the new scales, the low-string scale $M_{s}$ $\sim
10^{4}GeV$ and the singlet scalar VEV $\left \langle S\right
\rangle \sim
10^{3}GeV$. Then, we have investigated the possibility to consider this new
scalar as a viable DM connected to SM via the Higgs-mediated coupling $%
\lambda _{SH}$. Using the current constraints of the DM relic density $%
\Omega _{DM}\simeq 0.2$, we have approached the mass of the scalar DM $m_{S}$
(\ref{eq16}), bounded the range of the Higgs portal coupling $\lambda _{SH}$
(\ref{eq18}) and then we have dealt with the extreme cases.

This work comes up with many open questions related to DM problems using
string theory. In particular, a natural question is associated with the
physics of the QCD axion scalar field discussed in terms of the Peccei-Quinn
$U(1)_{PQ}$ symmetry corresponding to electroweak and supersymmetry breaking
scales in the context of closed and open string models. Moreover as argued,
the D6-brane physics finds naturally a place in M-theory compactifications
on G2 manifolds. It would be intersecting to understand such problems from
M-theory point of view.

\textbf{Acknowledgements}:  The authors would like to thank the
Instituto de Fisica teorica (IFT UAM-SCIC)  in Madrid for its
support via the Centro de Excelencia Severo Ochoa Program under
Grant SEV-2012-0249.  They   would like also to thank Maria Pilar
Garcia del Moral for discussions on related topics. AB  would like
to thank Luis Iba\~{n}ez for scientific supports.

\end{document}